\documentclass[useAMS,usenatbib,fleqn]{mn2e}
\usepackage{amsmath,amssymb,mathrsfs,graphicx,float,latexsym}
\usepackage{epstopdf,ragged2e}
\usepackage{hyperref}

\def\apj{{ApJ}}

\def\apjs{{The Astrophysical Journal Supplement}}
\def\apjl{{ApJL}}

\def\aap{{A\&A}}

\def\mnras{{MNRAS}}

\def\nat{{Nature}}

\def\prd{{Physical Review D}}
\def\pre{{Physical Review E}}

\def\04a{{2004 a}}
\def\04b{{2004 b}}

\title[No axially-isospectral neutron stars]{Astroseismology of neutron stars from gravitational waves in the limit of perfect measurement}
\author[A. G. Suvorov]{A. G. Suvorov\thanks{E-mail: suvorova@student.unimelb.edu.au}\\School of Physics, University of Melbourne, Parkville VIC
3010, Australia}
\begin{document}

\date{Accepted ?. Received ?; in original form ?}

\pagerange{\pageref{firstpage}--\pageref{lastpage}} \pubyear{?}

\maketitle

\label{firstpage}

\begin{abstract}

\noindent{The oscillation spectrum of a perturbed neutron star is intimately related to the physical properties of the star, such as the equation of state. Observing pulsating neutron stars therefore allows one to place constraints on these physical properties. However, it is not obvious exactly how much can be learnt from such measurements. If we observe for long enough, and precisely enough, is it possible to learn everything about the star? A classical result in the theory of spectral geometry states that one cannot uniquely `hear the shape of a drum'. More formally, it is known that an eigenfrequency spectrum may not uniquely correspond to a particular geometry; some `drums' may be indistinguishable from a normal-mode perspective. In contrast, we show that the drum result does not extend to perturbations of simple neutron stars within general relativity -- in the case of axial (toroidal) perturbations of static, perfect fluid stars, a quasi-normal mode spectrum uniquely corresponds to a stellar profile. We show in this paper that it is not possible for two neutron stars, with distinct fluid profiles, to oscillate in an identical manner. This result has the information-theoretic consequence that gravitational waves completely encode the properties of any given oscillating star: unique identifications are possible in the limit of perfect measurement.} 




\end{abstract}


\section{Introduction}

Studying pulsations of neutron stars allows us to get a glimpse into the properties of matter in extreme, astrophysical environments \citep{unno79}. For example, a young neutron star, formed due to a compact object merger or core collapse supernova, tends to oscillate and emit gravitational radiation (`ring-down') as it attempts to attain an equilibrium state \citep{nsring1,nsring2}. In general, fluid or local spacetime perturbations cause the host star to ring with a discrete set of oscillation frequencies for a certain period of time \citep{thorne1}. Because oscillation modes often couple to gravitational waves (GWs), the associated eigenfrequencies are complex \citep{vish70}. The real part gives the oscillation frequency of the mode, and the imaginary part gives the inverse of the damping time due to the emission of gravitational radiation \citep{schmidtrev}. These resonant frequencies can be studied through quasi-normal modes (QNMs), defined as the eigenmodes of oscillation, which can be used to describe the pulsations of any given star \citep{nollert99,sterg03}. The QNM spectrum is sensitive to the properties of the host star, such as the equation of state (EOS) \citep{e0s1,e0s2}, the magnetic field strength \citep{magfield1,magfield2}, or the presence of superfluidity \citep{superfluid1}. It is remarkable then that certain universal relations have been found to exist amongst QNMs for different kinds of neutron stars \citep{ilq1,ilq2}; though see \cite{ilq3}. 

\begin{figure}
\includegraphics[width=0.473\textwidth]{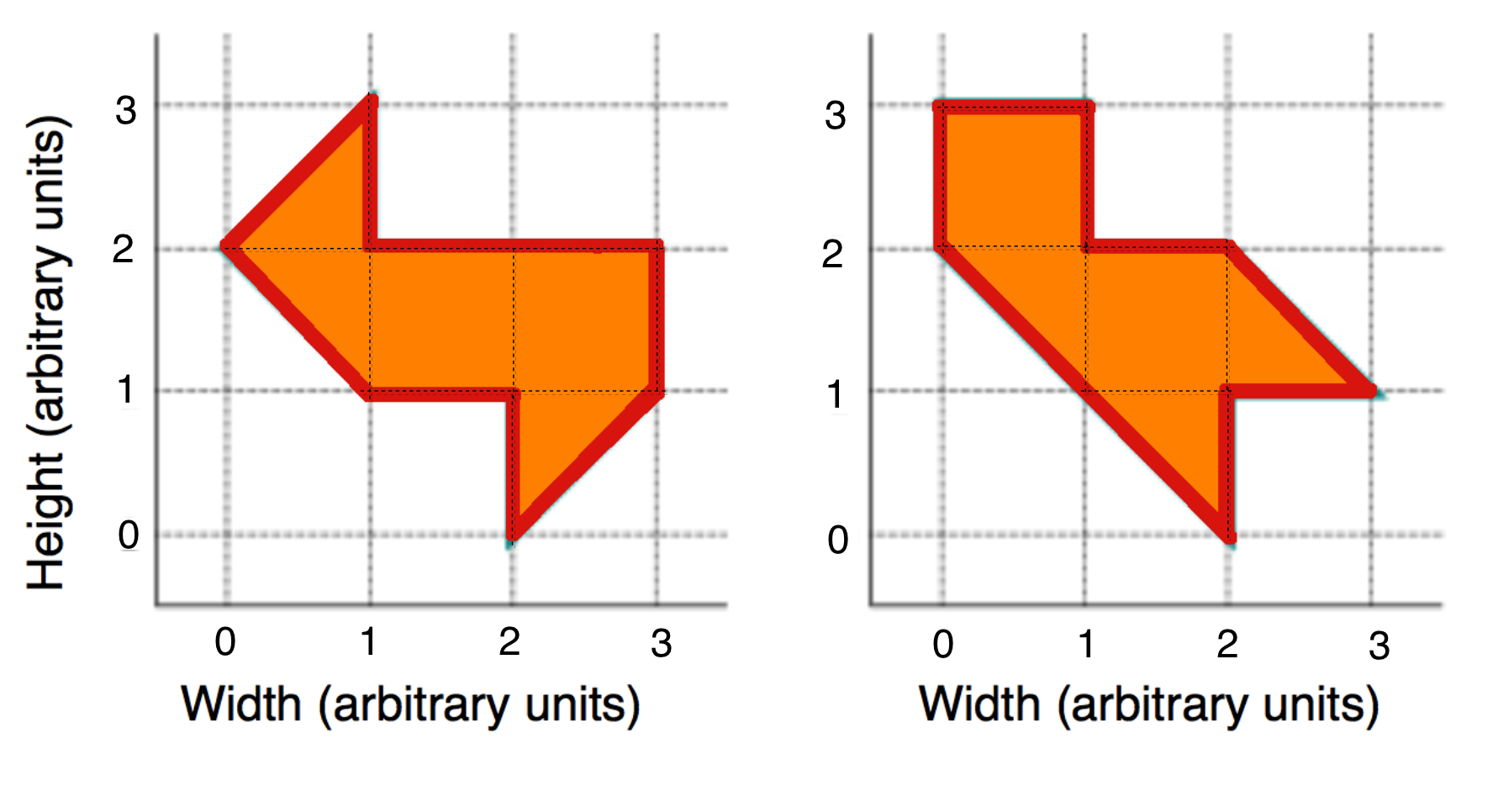}
\caption{An example pair of geometrical `drums' in two dimensions, first reported by \protect\cite{gordon}, which admit the same spectrum for the (Dirichlet) Laplace operator; they are isospectral. An observer listening only to sound waves produced by perturbing either of these drums would not be able to distinguish between them. \label{drums}}
\end{figure}

If there are in fact universal relations between QNMs of neutron stars, it is unclear precisely how much information can be gleaned from GW and other observations \citep{and96,kokkapo}. In the extreme case that two stars have the same spectrum, one faces a distinguishably problem of sorts. Similar issues arise in non-astrophysics contexts. One such problem is discussed by \cite{kac}, who poses the question of ``can one hear the shape of a drum?''. The problem effectively asks whether or not knowledge of the spectrum of the Laplace operator (`hearing' the frequencies) allows one to uniquely identify the geometry of a space (`drum'). The answer turns out to be no, meaning that one cannot always uniquely determine the geometry from the spectrum \citep{drum1}; several drums may be \emph{isospectral} \citep{drum2,gordon}. An example pair of isospectral drums in two dimensions are pictured in Fig. \ref{drums} \citep{gordon,gordon2}. {Furthermore, \cite{srid94} used differently shaped microwave cavities to experimentally demonstrate that isospectral domains exist.} The drum problem has appeared with various generalisations in the literature, such as considering drums on Riemannian manifolds \citep{riemdrum}, studying theoretical `fractal drums' to analyse Zeta functions \citep{zeta2,zeta1}, and hypothesising `quantum drums' whose spectra give insights into the nature of entanglement entropy \citep{ent4}. 



In this paper we consider a problem analogous to the one of \cite{kac}, but for oscillating neutron stars in general relativity: given the QNM spectrum of a ringing neutron star, can one uniquely determine its fluid properties? We show that, in contrast to the drum problem, the answer turns out to be yes for some simple models, meaning that two ringing neutron stars with different fluid properties are always distinguishable from a QNM perspective (Sec. 3). The implication of this result is that GW measurements can, in the limit of `perfect'\footnote{In this sense, \emph{perfect} means having a set of observations from which one can construct the full spectrum of oscillation eigenmodes.} measurement, uniquely identify all physical characteristics of any given perturbed star. We focus primarily on axial (sometimes called toroidal) perturbations of static, spherically symmetric stars because they are in many respects the simplest class (in a sense that is made precise in Sec. 2) of stellar pulsations \citep{chand75,wmode1}.  Some discussion on astrophysical implications is offered in Sec. 4. 

\section{Stellar structure}


In general, a static, spherically-symmetric compact object can be described through the line element\footnote{Throughout this work we adopt natural units with $G = c =1$, where $G$ is Newton's constant and $c$ is the speed of light. }

\begin{equation} \label{eq:sphlinel}
\begin{aligned}
ds^2 &= g_{\alpha \beta} dx^{\alpha} dx^{\beta} \\
&= -e^{2\nu} dt^2 + e^{2\lambda} dr^2 + r^2 d \theta^2 + r^2 \sin^2\theta d \phi^2,
\end{aligned}
\end{equation}
where $(t,r,\theta,\phi)$ are the usual Schwarzschild coordinates, and $\nu$ and $\lambda$ are functions of $r$ only. The Einstein equations 
\begin{equation} \label{eq:einstein}
G_{\mu \nu} = 8 \pi T_{\mu \nu},
\end{equation}
for the stress-energy tensor associated with a single, perfect fluid, viz. 
\begin{equation} \label{eq:stresstensor}
T^{\mu \nu} = \left( \rho + p \right) u^{\mu} u^{\nu} - p g^{\mu \nu},
\end{equation}
where $\rho$ is the energy-density, $p$ is the stellar pressure, $\boldsymbol{g}$ is the metric tensor defined in \eqref{eq:sphlinel}, and $\boldsymbol{u}$ is the 4-velocity of a generic fluid element, describe the structure of a static, non-rotating star \citep{tov1,tov2,unno79}. In particular, the metric function $\lambda$ is related to the mass distribution function $m(r)$, defined as the mass inside the circumferential radius $r$, through
\begin{equation}
e^{-2 \lambda} = 1 - \frac {2 m(r)} {r}.
\end{equation}
The functions $\nu(r)$, $\rho(r)$, and $p(r)$ are related through the equations of \cite{tov1,tov2}, which form the following system of differential equations
\begin{equation} \label{eq:tov1}
\frac {d \nu} { d r} = \frac {1} {p(r) + \rho(r)} \frac {d p} {d r},
\end{equation}
\begin{equation} \label{eq:tov2}
\frac {d m} {d r} = 4 \pi r^2 \rho(r),
\end{equation}
and
\begin{equation} \label{eq:tov3}
\frac {d p} {d r} = - \frac {\left[ \rho(r) + p(r) \right] \left[ m(r) + 4 \pi r^3 p(r) \right]} { r^2 \left[ 1 - \frac {2 m(r)} {r} \right]}.
\end{equation}
Supplemented by an EOS of the form $p = p(\rho)$, equations \eqref{eq:tov1}--\eqref{eq:tov3} uniquely determine the equilibrium stellar structure. The stellar radius $R_{\star}$ is defined by the vanishing of the stellar pressure, $p(R_{\star}) = 0$, and the total stellar mass $M_{\star}$ is given by $M_{\star} = m(R_{\star})$. Outside of the star $(r > R_{\star})$, where $p = \rho = 0$, the metric \eqref{eq:sphlinel} reduces to the Schwarzschild metric of mass $M_{\star}$ \citep{wald}.


\subsection{Axial perturbations}

In general, neutron star oscillations of small amplitude can be studied by simultaneously perturbing the Einstein equations \eqref{eq:einstein} and the stress-energy tensor \eqref{eq:stresstensor} given some background metric \eqref{eq:sphlinel} \citep{wald,schmidtrev}. One can introduce perturbations into quantities $q$ (e.g. $\rho, p, \boldsymbol{g}, \boldsymbol{u}$) by writing $q \rightarrow q^{(0)} + \delta q$, where $|\delta q|/ |q^{(0)}| \ll 1$ for equilibrium values $q^{(0)}$. Fourier-expanding each of the perturbed hydrodynamical and metric functions into angular and azimuthal harmonics of order $\ell,k$ [see e.g. \cite{ferrari08} for details] leads to a decoupled set\footnote{For rotating stars, the perturbation equations will, in general, not decouple in a simple way \citep{nondec2,nondec1}.} of differential equations, one governing \emph{polar} (poloidal) perturbations and the other governing \emph{axial} (toroidal) perturbations, i.e. the total spacetime metric tensor may be written as $g_{\mu \nu} = g_{\mu \nu}^{(0)} + \delta g_{\mu \nu}^{\text{axial}} + \delta g_{\mu \nu}^{\text{polar}}$, and the linearised Einstein equations for $\delta g_{\mu \nu}^{\text{axial}}$ and $\delta g_{\mu \nu}^{\text{polar}}$ decouple \citep{thorne1,moncrief,allen98}. The axial and polar modes are defined by how they transform under parity; under the transformation $\theta \rightarrow \pi - \theta$ and $\phi \rightarrow \pi + \phi$, axial modes of order $\ell, k$ transform as $(-1)^{\ell+1}$ while polar modes of the same order transform as $(-1)^{\ell}$ \citep{regwhel57,zerilli70}.

Polar perturbations correspond to regional compressions of the star (e.g. $f$-modes), whereas axial perturbation induce continuous differential rotations in the fluid (e.g. $r$-modes) \citep{wmode1,and96}. In this paper we concentrate on axial perturbations, since, as it turns out, axial oscillations do not modify the energy density or pressure of the fluid for static stars \citep{thorne1,thorne2}. Hence accounting for fluid back-reaction effects is trivial, which makes for a simple treatment of `axial-isospectrality' (see Sec. 3). 

After performing the aforementioned harmonic expansions, an axial perturbation of a static, spherically symmetric spacetime \eqref{eq:sphlinel} reduces to a single, Schr{\"o}dinger-like wave equation for functions $\tilde{Z}_{\ell}^{-}(t,r)$ defined as specific combinations of the tensor components of the metric perturbation $\delta g_{\mu \nu}^{\text{axial}}$ [see equation (20) of \cite{ferrari08}; similar equations arise for polar perturbations $\tilde{Z}_{\ell}^{+}(t,r)$]. Writing \begin{equation} \label{eq:qnmexpansion}
\tilde{Z}_{\ell}^{-}(t,r) = e^{i \omega t} Z^{-}_{\ell}(r),
\end{equation}
where $\omega$ is related to the angular velocity of the perturbed star and plays the role of the complex eigenfrequency of the system \citep{schmidtrev}, we obtain \citep{kokk94}

\begin{equation} \label{eq:perteqn}
\frac {d^2 Z^{-}_{\ell}} {dr_{\star}^2} + \left[ \omega^2 - V^{-}_{\ell}(r) \right] Z^{-}_{\ell} = 0,
\end{equation}
where $r_{\star}(r) = \int^{r}_{0} dr e^{\lambda - \nu}$ is the tortoise coordinate, and 
\begin{equation} \label{eq:potential}
V^{-}_{\ell}(r) = \frac {e^{2 \nu}} {r^3} \left\{ \ell \left( \ell +1 \right) r + 4 \pi r^3 \left[ \rho(r) + p(r)\right] - 6 m(r)  \right\}, 
\end{equation}
is the interior potential function. Outside of the star, the spacetime looks like a non-linear superposition of a Schwarzschild spacetime and GWs \citep{schmidtrev}, and the potential \eqref{eq:potential} reduces to the usual Regge-Wheeler form \citep{regwhel57},
\begin{equation} \label{eq:regwhel}
V^{-}_{\ell}(r) = \frac {1} {r^3} \left(1 - \frac{ 2 M_{\star}}{ r} \right) \left[ \ell \left( \ell + 1 \right) r - 6 M_{\star} \right].
\end{equation}

The axial QNM spectrum is therefore determined by solving equation \eqref{eq:perteqn}, using continuous `total' potential functions $V_{\ell}(r)$  given by \eqref{eq:potential} inside the star $(r \leq R_{\star})$ and \eqref{eq:regwhel} outside of the star $(r > R_{\star})$. The boundary conditions on \eqref{eq:perteqn} are chosen such that we have pure outgoing radiation at infinity, i.e. that at  $r_{\star} \rightarrow \infty$ we have \citep{chand75,kokk94}
\begin{equation} \label{eq:atinfinity}
Z^{-}_{\ell}(r) \sim e^{-i \omega r_{\star}(r)},
\end{equation}
while at the center of the star $(r=0)$ the perturbation functions $Z^{-}_{\ell}$ are assumed to be regular,
\begin{equation} \label{eq:atcentre}
Z^{-}_{\ell}(r) \sim r^{\ell +1}.
\end{equation}
Equations \eqref{eq:perteqn}--\eqref{eq:regwhel}, subject to the boundary conditions \eqref{eq:atinfinity} and \eqref{eq:atcentre}, constitute an eigenvalue problem \citep{thorne1,thorne2,nondec2}.

Since we are concerned with certain uniqueness properties of axial perturbations (Sec. 3), it is important to discuss whether an expansion of the form \eqref{eq:qnmexpansion} is always permitted over the background \eqref{eq:sphlinel}. Spacetimes which allow for such a decomposition are said to be \emph{complete} with respect to QNMs; in this context, completeness means that it is possible to express the evolution of a wavefunction as a sum over eigenfunctions \citep{price92,nollert99}. In fact, this is largely still an open problem within general relativity \citep{and93,ching95,beyer99,pal15}. For example, it has been argued that if the potential $V$ has a significant tail at large radii, the GWs emitted due to the perturbation may only have power-law decays at late times, which means that the signal cannot be represented through QNMs, which decay exponentially \eqref{eq:atinfinity} \citep{price92,ching96}. However, \cite{ching96} developed a set of criteria (e.g. the integral $\int_{0}^{\infty} d r_{\star} r_{\star} |V|$ is finite) which, if satisfied for a given spacetime [e.g. \eqref{eq:sphlinel}], imply that the QNM spectrum is complete; see also \cite{newt60,chand75,ho98}. We expect the conditions of \cite{ching96} to hold for any physically reasonable star \citep{chand75,price92}, though our analysis would need to be performed at the GW level (see Sec. 2.2) if this were not the case since we cannot apply \eqref{eq:qnmexpansion}.


\subsection{Gravitational waves}

While axial perturbations of neutron stars do not modify the stellar pressure or density, a time varying quadrupole moment is generated by the inducement of a continuous, non-varying differential rotation \citep{kokk94,and96}. This leads to the release of gravitational radiation \citep{thorne2,thorne80}. The properties of the emitted GWs are encapsulated in the QNM spectrum, which is used to describe the properties of $\delta g_{\mu \nu}^{\text{axial}}$ \citep{tominaga,detweiler}. 

The plus $h^{+}$ and cross $h^{\times}$ GW polarisations of the perturbed system are given by \citep{thorne2,tominaga} [see also equations (1) through (5) of \cite{ferrari08}]
\begin{equation} \label{eq:plus}
h^{+}(t,r,\theta,\phi) = -\frac {1} {2 \pi} \int d \omega \frac {e^{i \omega \left( t - r_{\star}\right)}} {r} \sum_{\ell m} \left[ \frac {Z^{-}_{\ell m}(r, \omega)} {i \omega} \frac {X^{\ell m}(\theta,\phi)} {\sin \theta} \right],
\end{equation}
and
\begin{equation} \label{eq:times}
h^{\times}(t,r,\theta,\phi) = \frac {1} {2 \pi} \int d \omega \frac {e^{i \omega \left( t - r_{\star}\right)}} {r} \sum_{\ell m} \left[ \frac {Z^{-}_{\ell m}(r, \omega)} {i \omega} \frac {W^{\ell m}(\theta,\phi)} {\sin \theta} \right],
\end{equation}
respectively, where
\begin{equation} \label{eq:xharmonic}
X^{\ell m}(\theta,\phi) = 2 \left( \frac {\partial^2} {\partial \theta \partial \phi} - \cot \theta \frac {\partial} {\partial \phi}  \right) Y^{\ell m}(\theta,\phi),
\end{equation}
and
\begin{equation} \label{eq:wharmonic}
W^{\ell m}(\theta,\phi) = \left( \frac {\partial^2} {\partial \theta^2} - \cot \theta \frac {\partial} {\partial \theta} - \csc^2\theta \frac {\partial^2} {\partial \phi^2} \right) Y^{\ell m}(\theta,\phi),
\end{equation}
are angular functions defined in terms of the spherical harmonics $Y^{\ell m}(\theta, \phi)$.

Some combination of expressions \eqref{eq:plus} and \eqref{eq:times} can, in principle, be observed with an interferometer such as the Laser Interferometer Gravitational-Wave Observatory (LIGO) \citep{jaran98}, which then allows one to determine the stellar properties. For our purposes, the important point is the following: in a QNM-complete spacetime, a given potential $V^{-}_{\ell}$ uniquely [due to standard theorems in Sturm-Liouville theory \citep{sturm1}] determines a function $Z^{-}_{\ell}$ from \eqref{eq:perteqn}, which, through expressions \eqref{eq:plus} and \eqref{eq:times}, determines the resulting GW signal. GW analysis can therefore allow us to, in the limit of perfect measurement, uniquely reconstruct $V^{-}_{\ell}$ experimentally, in-turn uniquely identifying all stellar properties from \eqref{eq:potential} if no two stars can ever be axially-isospectral.

\section{Axial-isospectrality}

The key ingredient for determining the properties of axial perturbations are the variables $Z_{\ell}^{-}$, which solve the scalar eigenvalue problem \eqref{eq:perteqn} described in Sec. 2.1. However, the $Z_{\ell}^{-}$ are themselves uniquely determined by the potential functions \eqref{eq:potential} and \eqref{eq:regwhel}, which depend on the metric and hydrodynamical variables, which are in-turn constrained by the Tolman-Oppenheimer-Volkoff (TOV) equations. In fact, the structure of a static, spherically symmetric, perfect fluid neutron star is ultimately determined by three functions (noting that $\lambda$, $m$, and $\rho$ are not independent): $\nu, m,$ and $p$. We say that a pair of stars are `axially-isospectral' if their associated axial potentials $V_{\ell}^{-}$ \eqref{eq:potential} and Regge-Wheeler potentials \eqref{eq:regwhel} are identical; such stars will have isomorphic axial QNM spectra.

\subsection{No axially-isospectral stars}

{Here we show that there does not exist distinct solutions to the TOV equations \eqref{eq:tov1}--\eqref{eq:tov3} which also satisfy the axial-isospectrality conditions, i.e. which have matching interior \eqref{eq:potential} and exterior \eqref{eq:regwhel} potentials. To this end, we show that a given total potential $V_{\ell}$ uniquely corresponds to a set of metric and hydrodynamical variables. This implies that distinct stars with the same functions $V_{\ell}$ cannot exist, which allows us to conclude that non-trivially axially-isospectral, static, perfect fluid stars cannot exist.}

{From expression \eqref{eq:potential}, we see that $V_{\ell}$ can be decomposed into a part which is dependent on $\ell$ and a part which is independent from $\ell$. Given $V_{\ell}$, one can therefore identify both of these two components independently. The key point is now that these two pieces of information together uniquely determine an EOS, thereby yielding a unique solution to the TOV equations \eqref{eq:tov1}--\eqref{eq:tov3} \citep{shap82}. }

{To see this explicitly, suppose we are `given' some (set of) $V_{\ell}$, which, inside the star, can be deconstructed as (say) $V_{\ell}^{-}(r) = A(r) + \ell \left(\ell + 1\right) B(r)$, where $A$ and $B$ are known functions which do not depend on $\ell$. Under these circumstances, we may treat \eqref{eq:tov1} as a first-order, linear, inhomogeneous differential equation for $p$, viz.
\begin{equation} \label{eq:pint2}
0 = \frac {d p} {d r} - \left[ p(r) + \rho(r) \right] \frac {2 B(r) + r \frac {d B}{ d r}} {2 r B(r)},
\end{equation}
where we have made use of the relation $B(r) = e^{2\nu(r)}/r^2 $ from \eqref{eq:potential}. Expression \eqref{eq:pint2} alone is not enough to determine the EOS because the relationship between $\rho$ and $r$ is unconstrained. However, the term $m(r)$ can be expressed in terms of $p$, $\rho$, and the known functions $A$ and $B$. As such, equation \eqref{eq:tov2} becomes
\begin{equation} \label{eq:pint3}
\begin{aligned}
0 =& 4 \pi r^2 B(r) \left\{ 3 \left[ p(r) - \rho(r)\right] + r \left( \frac {d p} {d r} + \frac {d \rho} { dr} \right) \right\}  \\
&+ A(r) \left[\frac {2 B(r) + r \frac {d B}{ d r}} {B(r)} - 3 \right] - r \frac {d A} {d r},
\end{aligned}
\end{equation}
where we have made the identification $A(r) = 4 \pi e^{2 \nu(r)} \left[ \rho(r) + p(r) \right] - 6 e^{2\nu(r)} m(r)/r^3$.
Equation \eqref{eq:pint3} acts as a second, linear, first-order differential equation for the variables $p$ and $\rho$.  Combining the above, we have that knowledge of $V_{\ell}$ translates into two linear equations in two unknowns $\rho$ and $p$, i.e. \eqref{eq:pint2} and \eqref{eq:pint3}. Note that \eqref{eq:pint2} and \eqref{eq:pint3} must necessarily be consistent with \eqref{eq:tov3}, else the given functions $V_{\ell}$ could not correspond to \emph{any} star. }

{The boundary conditions for \eqref{eq:pint2} and \eqref{eq:pint3} are the standard ones applied to the TOV equations, namely \citep{shap82}
\begin{equation} \label{eq:bc1}
m(0) = 0,
\end{equation}
and
\begin{equation} \label{eq:bc2}
\nu(R_{\star}) = \frac {1} {2} \ln \left( 1 - \frac {2 M_{\star}} {R_{\star}} \right).
\end{equation}
The former condition ensures that the circumferential mass $m(r)$ is well defined at the origin and acts as an initial condition for $\rho$, while the latter is the Schwarzschild matching condition which, although satisfied by $\nu(r)$, necessarily constrains the relationship between $\rho$ and $p$ near the boundary through \eqref{eq:pint2} by noting that $p(R_{\star}) = 0$ and $m(R_{\star}) = M_{\star}$  by definition. One therefore has a well posed system and can, in principle, solve for $p$ and $\rho$ in terms of the known functions $A$ and $B$ uniquely due to standard theorems on ordinary differential equations \citep{sturm1}.}

{In summary, given $V_{\ell}$, one can uniquely identify $\nu$, $p$, and $\rho$ through the TOV equations \eqref{eq:pint2} and \eqref{eq:pint3} subjected to the standard boundary conditions \eqref{eq:bc1} and \eqref{eq:bc2}. As noted in Sec. 3, this implies that the stellar structure is completely determined from $V_{\ell}$, and we can therefore conclude that no two stars can admit the same axial QNM spectrum.}




\section{Discussion}

In this paper, inspired by the drum problem popularised by \cite{kac}, we explore the possibility that neutron stars, with distinct fluid profiles, might exhibit the same QNM spectrum and thereby `ring' in an identical way. It is known, for example, that two stars which have different magnetic field topologies can admit the same mass quadrupole moments \citep{mastr1}. We consider the case of axial perturbations and show that `axially-isospectral' neutron stars cannot exist. This result suggests that one can, in principle, uniquely determine the nuclear EOS (and all other observables of interest) from GW and QNM measurements\footnote{Assuming that the spacetime is QNM-complete in the sense discussed in Sec. 2.1.}.  From an information-theoretic standpoint, there is a one-to-one relationship between `perfect' GW measurements and stellar structure. This is in contrast to the drum problem, where it is known that it may be impossible to uniquely determine the geometry from the normal mode spectrum \citep{drum2,gordon,drum1}; see also Fig. \ref{drums}.


Eventually, a measurement of continuous GWs from a ringing neutron star, using e.g. LIGO \citep{ligo1}, will provide insights into the behaviour of nuclear matter at very high $(\sim 10^{18} \text{ kg m}^{-3})$ densities \citep{unno79,and96}. In this paper we consider axial perturbations, which generate a time varying quadrupole moment by inducing a continuous, non-varying differential rotation \citep{thorne1,kokk94}. Axial modes of rotating neutron stars are known to be prone to various instabilities wherein the canonical energy becomes negative, such as the instability discussed by \cite{fm98} [see also \cite{chand70,fs78}]. These instabilities can cause axial (e.g. $r-$) mode amplitudes to grow [see \cite{hask15} for a review], which could allow for most of the rotational energy and angular momentum of the star to be carried away by the GWs \citep{fm98,koj99}. There is a well-known observational discrepancy that all known neutron stars spin much slower than the break-up limit, which is particularly puzzling for low-mass X-ray binaries (LMXBs) where one expects accretion to significantly spinup the star \citep{chak03}. Depending on the importance of other dissipative processes that may compete with GW emission in removing energy from the star (e.g. neutrino diffusion or viscosity), low LMXB spins may potentially be explained by unstable axial modes \citep{bildstenpaper}. If axial mode amplitudes are large enough to account for the discrepancy, it is possible that the associated GWs will be detected in the near future. While the result presented here cannot be used to assist in detection efforts, and applies only to the idealised situation of `perfect' measurement, it implies that GW spectra provide complete information about their hosts which further supports the importance of GW astronomy.


Aside from (axial-)isospectrality, one might ask whether two stars could be `nearly-isospectral' in some appropriate sense. In reality, a combination of instrumental [e.g. thermal noise \citep{lev98}] and systematic [e.g. use of response functions \citep{respfun}] error prevents one from ever perfectly knowing a QNM spectrum. There may also be issues of sampling since, for example, the Nyquist bound \citep{nyquist} prevents an exact waveform identification for all $\omega$ from discrete measurements. Therefore, if two stars were to admit QNM spectra which deviate, in some as-of-yet undefined way, by a small, non-zero amount, they may still be experimentally indistinguishable \citep{bai17}. These two stars would appear as effectively isospectral, even if there is not an exact mapping between the two spectra. Questions of this sort have interesting implications about the information content of GW signals. The details of `near-isospectrality' will be investigated in future work. It would also be worthwhile considering whether isospectral stars in modified theories of gravity can exist, since the properties of GWs can vary \citep{gwmod1,gwmod2}. The solution generating techniques discussed by \cite{suvmel2} may be useful in this direction.




%

%


\section{Acknowledgements}
We thank Prof. Bill Moran for discussions. {We thank the anonymous referee for providing useful feedback which improved the quality of the manuscript.}

\bsp \label{lastpage}

\end{document}